# Superconducting properties and *c*-axis superstructure of $Mg_{1-x}Al_xB_2$

J. Y. Xiang, D. N. Zheng*, J. Q. Li, L. Li, P. L. Lang, H. Chen, C. Dong, G. C. Che,
Z. A. Ren, H. H. Qi[#], H. Y. Tian, Y. M. Ni and Z. X. Zhao
*National Laboratory for Superconductivity, Institute of Physics and the Center for Condensed Matter Physics*
*Chinese Academy of Sciences, Beijing 100080, People's Republic of China*
[#]*also in College of Materials science and Chemical Engeering, Yanshan University*
*Qinghuangdao 066004, People's Republic of China*

## Abstract

The superconducting and structural properties of a series of $Mg_{1-x}Al_xB_2$ samples have been investigated. X-ray diffraction results confirmed the existence of a structural transition associated with the significant change in inter-boron layer distance as reported previously by Slusky *et al*. Moreover, transmission-electron-microscopy observations revealed the existence of a superstructure with doubled lattice constant along the *c*-axis direction. We propose that this superstructure is probably related to the structural transition. The modifications of superconducting transition temperature $T_c$, the normal state resistivity, and the upper critical field $B_{c2}(0)$ by Al doping are discussed in terms of Al-substitution induced changes in the electronic structure at the Fermi energy.

## I. INTRODUCTION

The discovery of superconductivity in magnesium diboride with a transition temperature $T_c$ of 39 K has attracted considerable interest.[1] This is because this metallic binary compound, which has been known since 1953,[2] consists of no transition metals and yet exhibits a remarkably high transition temperature, higher than the value obtained in any other metallic binary compounds. $MgB_2$ adopts a very simple hexagonal crystal structure, comprising interleaved two-dimensional boron and magnesium layers.[2] The appearance of superconductivity in $MgB_2$ with such a high value of $T_c$ immediately raises the speculation for even higher superconducting temperatures in conventional metallic binary materials. On the other hand, the underlying mechanism of superconductivity in this compound is still an issue of current debate. There are mainly two competitive theories of superconductivity in $MgB_2$. The first one is based on the well-established phonon-mediated BCS theory, [3] and the high $T_c$ value is believed to be due to the high phonon frequencies and strong electron-phonon interactions. This theory is supported by the results of a number of different experiments such as isotope effect, [4] quasi-particle tunneling, [5, 6] specific heat, [7] photoemission spectroscopy [8] and inelastic neutron scattering.[9] The second theory concerning the superconductivity in $MgB_2$ was put forward by Hirsch[10, 11] who proposed a 'universal' mechanism where superconductivity in $MgB_2$ was driven by the paring of dressed holes. Indeed, the hole character of the carriers in $MgB_2$ was confirmed by Hall measurements. [12] Soon after the discovery of superconductivity in $MgB_2$, Slusky *et al*.[13] reported studies of how the behavior responds to the substitution of Al for Mg in this compound. They observed that $T_c$ decreases smoothly by a few degrees with increasing *x* for $0<x<0.1$ and bulk superconductivity disappears completely due to a structural instability for $x>0.1$. In this paper, the effect of partial Al substitution for Mg on the structural and superconducting properties of $MgB_2$ is studied. In particular we have, for the first time, observed a superstructure in Al doped samples. We suggest that this superstructure is related to the structural transition reported by Slusky *et al.* [13] The superconducting transition temperature, normal state resistivity and the upper critical field were also studied. The variation of these parameters with Al substitution concentration is discussed in connection with the change in the structure and the density of states at the Fermi energy.

## II. EXPERIMENTAL

Polycrystalline samples used in this study were prepared by a conventional solid reaction process. Two series of samples with similar fabrication process were made. Starting materials were elemental powders of Mg, Al and B. The raw materials with nominal composition of $Mg_{1-x}Al_xB_2$ were well mixed and pressed into discs, with 8 mm in diameter and 0.4 gram in weight. These discs were wrapped with Ta foil and sealed in quartz tubes under a background pressure about $10^{-3}$ Pa. The sealed quartz tubes were placed in a tube furnace, and the temperature was ramped at a rate of 100 °C per hour to 900 °C and held for two or three hours. This was followed by furnace cooling to the room temperature.

The microstructures and phase purity of the samples were studied by means of X-ray diffraction (XRD) and transmission electron microscopy (TEM). For TEM examination, wafers of the samples were first metallographically polished and then thinned by ion milling. Local composition was probed by energy-dispersive X-ray spectroscopy (EDX).

Resistivity measurements were made on bar-shaped samples of approximate dimensions $6\times3\times2$ mm using a four-probe technique with a current of 10 mA. AC-susceptibility measurements were also performed to determine the superconducting transition temperature. DC-magnetization was measured in a Quantum Design SQUID magnetometer.

## III. RESULTS AND DISCUSSION

The X-ray powder diffraction patterns obtained on the samples showed the predominant phase was of $MgB_2$ type and there was some minor amount of impurity phases such as MgO presented in the samples. The partial representation of the XRD patterns in the $2\theta$ range of 50 to 60 degree is shown in Fig. 1. The results are similar to that reported by Slusky *et al.*, [13] *i.e.*, the substitution of Al results in a significant change in the *c*-axis lattice parameter while the in-plane lattice parameter is relatively constant. In the range approximately between $x=0.09$ to $x=0.25$, the (*002*) reflection peak becomes broad, indicating the partial collapse of the separation between boron sheets and the presence of two isostructure phases. Fig. 2 shows the lattice parameters as a function of Al substitution level. In the shadowed region, because of the peak broadening, there is a large uncertainty associated with the value of *c*. Nevertheless, the data in Fig.2 show that in the shadowed region there appears a discontinuity, which could be related to a structural transition as pointed out in Ref .13.

Resistivity measurements were carried out on two series of samples that were fabricated under almost the same conditions (one was sintered at 900°C for 2 hours and another one was at the same temperature for 3 hours). The intention was to investigate how the normal state resistivity varied with Al doping concentration. However, the data obtained did not show a clear trend of the variation for the resistivity with Al content over the whole doping range we investigated ($0<x<0.6$). In particular, there was no sharp rise in the resistivity around the phase transition around $x=0.1$ as what would be observed in alloys of metals. There are a number of factors that could mask the intrinsic resistivity of the materials studied. These include sample density variation, grain boundary conditions, and disorder scattering. For the all samples measured, we checked the density and found that the values were around 70 per cent of the theoretical ones while the variation between the samples was within 10 percent. Thus, the porosity of the samples would certainly affect the absolute resistivity values, but its effect on the relative variation of the resistivity might not be substantial. The resistivity is around 30

$\mu\Omega$.cm at 40 K for undoped samples.

Although the resistivity data were rather scatter and showed no clear trend with the increase of the Al level over the whole doping range, we found that in the low doping region ($x<0.1$) the normal state resistivity increases monotonically for the both series of samples we made. This is in agreement with the observation of Lorenz et al. [14]. The increase of the normal state resistivity (at least within the doping level of 0.1) by Al doping is consistent with the hypothesis that the charge carrier in MgB$_2$ is hole-like [10, 11] since the substitution of Al$^{3+}$ for Mg$^{2+}$ reduces carrier density and subsequently leads to an increase in the normal state resistivity. It has been suggested that the conduction in MgB$_2$ occur primarily in the B layer [3] and thus the effect of Al doping on the Mg site on the impurity scattering is small while the effect of charge transfer is more significant. The results reported in Ref.14 also showed that an increase in thermopowers in Al doped samples, indicating the reduction of carrier numbers.

In Fig. 3, we illustrate the evolution of superconducting transition with Al doping in Mg$_{1-x}$Al$_x$B$_2$ by presenting a series of zero-field-cooled magnetization versus temperature curves measured on the samples with a variety of different doping levels. The measurements were carried out on bar-shaped samples. At the low doping level, the transition is sharp and the transition decreases slightly with the increase of the Al doping level. As the Al concentration increases, the transition becomes broad and there appear two transitions as indicated in the graph. We believe that the first transition is the transition within grains while the second one is associated with the establishment of bulk screening currents. For the doping levels lower than $x=0.3$, the full diamagnetism signal is obtained whereas at higher doping levels the bulk superconductivity disappears. Furthermore, in the high doping region ($x>0.3$), not only the bulk superconductivity disappears, but also the diamagnetism magnetization becomes smaller and smaller. This is attributed to the decreased superconducting volume fraction within grains in the Al doped samples resulted from the appearance of a non-superconducting phase.[13]

In Fig. 4, we show the variation of superconducting transition temperature as a function of Al substitution level. In the graph we present the data of both onset superconducting transition temperature $T_{c(onset)}$ and the bulk transition temperature $T_{c(bulk)}$. The determination of these two characteristic temperatures is illustrated in the inset of Fig. 3.

Because the magnetization data in the temperature range between $T_{c(onset)}$ and $T_{c(bulk)}$ show rounded curvature, there is a large uncertainty associated with $T_{c(onset)}$. Therefore the value in Fig. 3 (open circles) might regarded as the lower bound of $T_{c(onset)}$. The dashed line in the graph represents the upper bound. For the samples with $x>0.4$, there is still a very small but measureable diamagnetism signal. In the low doping region the two transition temperatures are close and decrease slowly with increasing Al doping level. For the doping levels above $x=0.1$, $T_c(bulk)$ drops more quickly than $T_c(onset)$. The $x=0.1$ composition coincides approximately with the substitution level at which two-phase region appears. As pointed out by Slusky et al.,[13] the compound MgB$_2$ is near a structural instability, at slightly higher electron concentration, that can destroy the superconductivity. Their experimental results and the results presented here suggested that the suppression of bulk superconductivity occurs at the same Al doping level at which a structural transition occurs. After the structural transition, a presumably non-superconducting phase appears, and the two phases are intergrown on a nanometer length scale. Hence the granular nature of the superconducting phase and proximate effects, caused by the intergrowth, together with the possible non-uniform distribution of Al contents are expected to influence the shapes of the curves in Fig. 3 and the $T_c$ dependence in Fig. 4.

On the other hand, the decrease of $T_c$ at substitution levels lower than $x=0.1$ has been attributed to a density of states effect. [13, 14] In MgB$_2$ the Fermi energy $E_F$ is close to an edge of rapidly decreasing density of states and any increase of $E_F$ by Al substitution could result in a significant decrease of the density of states at Fermi energy $N(E_F)$, which, in the BCS formulism, leads to a decrease of $Tc$. In the next few paragraphs, we present the results of the upper critical field $B_{c2}(0)$ and try to deduce some information concerning $N(E_F)$.

In the inset of Fig. 5 magnetic moment versus temperature curve measured at 0.5 T for the $x=0.07$ sample is shown. A small normal state background contribution has been subtracted from the data. Like commonly observed in high-$Tc$ superconductors, there were two characteristic temperatures $Tc$ and $T_{irr}$. In the region between these two temperatures, field-cooled (FC) and zero-field-cooled (ZFC) magnetization data were overlapped. The superconducting transition temperature was defined as the intercept of a linear extrapolation of the magnetic moment in the superconducting state with the normal state base line. The field dependence of $Tc$ gives the upper critical field $B_{c2}(0)$ data, which are presented in Fig. 5 for the two samples with $x=0$ and $x=0.07$, respectively. For both samples, $B_{c2}(0)$ shows approximately linear temperature dependence with a slope of -0.6 T/K and -0.58 T/K, respectively. For the $x=0$ sample, the value is similar to that reported in the literature. [15]

The relation between the zero-temperature upper critical field $Bc2(0)$ and $dBc2/dT$ at $Tc$ is: [16, 17]

$$B_{c2}(0) = 0.5758 \, [\kappa_1(0)/\kappa]_{Tc} \, (dHc2/dT)|_{Tc} \quad (1)$$

In the dirty limit $\kappa_1(0)/\kappa = 1.20$ [18, 19] while in the clean limit $\kappa_1(0)/\kappa = 1.26$.[20] These expressions yields $Bc2(0)=16.2$ T for $x=0$ and $Bc2(0)=14.8$ T for $x=0.07$, respectively, assuming the dirty limit. Similarly, we obtain $Bc2(0)=16.8$ T for $x=0$ and $Bc2(0)=15.4$ T for $x=0.07$, respectively, assuming the clean limit. Clearly, the decrease of $Bc2(0)$ is primarily due to the decrease of $Tc$ and the slope is barely changed. The coherent length at zero-temperature $\xi(0)$ may be estimated using the expression $Bc2(0) = \Phi_0/(2\pi\xi^2(0))$. For the undoped $MgB_2$ sample the value is $\xi(0)=4.5$ nm in the dirty limit and $\xi(0)=4.4$ nm in the clean limit.

The coherence length we just calculated is the GL coherence length. In order to know whether the samples are in the clean limit or in the dirty limit, we need to compare the BCS coherence length $\xi_0$ and the electronic mean-free-path $l$. The former can be estimated using the relation [21] $\xi_0 = 0.15\hbar v_F/k_B Tc$, where $v_F$ is the Fermi velocity. Taking an average value [3] of $v_F = 4.7 \times 10^7$ cm/s, we obtain $\xi_0=14$ nm. On the other hand, for estimating $l$, we use the carrier density $6.7 \times 10^{22} cm^{-3}$ (2 carriers per unit cell)[22] and the resistivity value of 30 $\mu\Omega$.cm for our undoped $MgB_2$ sample. Then, $l$ is estimated to be approximately 1 nm. This clearly puts the sample in the dirty limit. However, since the sample porosity and other extrinsic factors would certainly result in overestimated resistivity values, we should be cautious with the value of $l$. Canfield et al.[22] reported very low resistivity values on dense $MgB_2$ wires, and subsequently a long mean-free-path and the clean limit behavior was inferred. The resistivity value reported in the literature varies greatly for MgB2. We note that even for high quality epitaxial thin film samples resistivity values [23, 24] are about one order of magnitude higher than those reported in Ref.22. In particular, the recent work by Eom et al.[25] and Patnaik et al.[26] has shown that oxygen could be alloyed onto B site which leads to a significant increase in resistivity and the upper critical field. Therefore, it is plausible that with increasing sample resistivity $MgB_2$ may change from the clean limit to the dirtily limit. Furthermore, the data obtained on our $MgB_2$ samples show that the GL coherence length $\xi(0)$ is substantially shorter than the BCS coherence length $\xi_0$, an indicative of dirty limit behavior.

In assuming the dirty limit, we may infer some information about the density of states from the upper critical field data. As discussed above, the band calculation showed that in $MgB_2$ the Fermi energy is situated around a peak in the density of states. [3, 27, 28] The substitution of Al increases the electron transfer from the Mg plane. In a rigid band picture, this would mean a decrease of the density of states at the Fermi energy $N(E_F)$. In the dirty limit, $N(E_F)$ can be related to the slope of $Bc2$ near $Tc$ with the relation $dB_{c2}/dT|_{Tc} \sim \rho N(E_F)$, where $\rho$ is the residual normal state resistivity. Clearly, the slight decrease of $dBc2/dT$ and the increase in the normal state resistivity indicate that $N(E_F)$ is indeed reduced in the Al doped sample. If the resistivity could be determined more accurately, a more quantitative estimation of $N(E_F)$ could be made.

In conventional alloy and intermetallic-compound superconductors such as NbTi and $(Nb,Ti,Ta)_3Sn$, the upper critical field $Bc2(0)$ can be increased by substitutional alloying which results in resistivity increases while has a much mild effect on $Tc$ and density of states. This can be illustrated by the direct dependence of $B_{c2}(0)$ on the normal state resistivity through the formula [29] $Bc2(0)=3.110\rho\gamma Tc$ (in SI units). Here $\gamma$ is the electronic specific heat coefficient. The work by Eom *et al.* [25] and Patnaik *et al.*[26] appears to suggest that oxygen could be alloyed into MgB2 on the B site and lead to a dramatic increase in the upper critical field even though $Tc$ is reduced slightly.[25] In contrast the data shown here indicate that although low-level Al doping can increase the normal state resistivity substantially in $MgB_2$ both $Bc2(0)$ and $Tc$ are reduced. This is direct evidence that the rise in $\rho$ is compensated by the decrease in the density of states.

In order to characterize the structural features associated with the structural transition as reported in the present system, we have carried out TEM investigations on the Al doped samples, especially, with $x$ ranging from 0.1 to 0.25. As a result, a variety of structural phenomena have been revealed in these materials, such as phase separation and intergrowth of two structural phases. The most striking structural feature found in this investigation is the presence of a superstructure phase of doubled lattice parameter along the $c$-axis direction. Fig.6 presents the electron diffraction patterns for two samples taken along the $b$-axis direction. Fig. 6a is the electron diffraction pattern of the pure $MgB_2$ sample, all diffraction spots can be well indexed by the hexagonal cell with lattice parameters of $a$=3.07Å and $c$=3.57Å. The pattern in Fig. 6b was taken on the $x$=0.25 sample. In addition to the basic reflection spots, evident superstructure spots can be clearly recognized at the systematic ($h, k, l$+1/2) positions. TEM examination also revealed that the grains were very small, on the order of 0.1μ in the $x$=0.25 sample. In Fig. 7, we show a dark-field image of a grain for the $x$=0.25 sample. It would be desirable to fabricate samples with larger grains that would allow more detailed TEM study. Work along this line is undergoing. Nevertheless, we could see that the phase with the superstructure intergrows with the normal $MgB_2$ phase.

The superstructure, however, was not observed in XRD spectra. In principle, the appearance of the superstructure with doubled $c$-axis parameter would result in an additional (*00l*) peak at low $2\theta$ angle. The XRD data showed no sign of such a peak within the instrumental sensitivity. We propose that this is due to the very similar X-ray scattering factor for Mg and Al atoms. The diffraction of X-ray by atoms in a crystal lattice is due to the interaction between the surrounding electrons and X-ray. Mg and Al are elements that occupy positions next to each other in the periodic table, and the electron cloud distribution is very similar for $Mg^{2+}$ and $Al^{3+}$. Therefore the scattering factor is almost the same for the two elements. Furthermore, the phase with superstructure is presented in small regions, especially for the low Al doped samples, and hence the XRD peak may become smearing and difficult to be observed. On the other hand, TEM can probe the structure in a comparatively small region. Moreover, the diffraction of charged electrons is more sensitive to the electric field potential of atoms that are different for Mg and Al atoms. Thus, the superstructure can be observed more readily by TEM.

In order to ensure that the observed superstructure is resulted from Al substitution, we made a systematical examination on two sets of samples with 0< $x$<1. The results indicate that this superstructure appears in a large range with 0.1< $x$< 0.7 together with both Al ordering and the resultant structural distortion. A more detailed analysis about atomic structure in this superstructure phase will be reported in a forthcoming paper. [30]

In summary, we have investigated the effect of Al substitution for Mg on the structural and superconducting properties of $Mg_{1-x}Al_xB_2$ samples. The X-ray diffraction results confirmed the existence of a structural transition associated with the significant change in inter-boron layer distance as reported previously by Slusky *et al*.[13] Moreover, transmission electron microscopy examination showed the existence of a superstructure in the direction perpendicular to the boron honeycomb layers with doubled *c*-axis lattice parameter. We propose that the observed superstructure be related to the structural transition. The experimental results showed that the normal state resistivity was increased and *Tc* was decreased with increasing the Al concentration, respectively. The upper critical field $Bc2(0)$ was determined for samples with low Al concentrations and found to be reduced by Al doping. Furthermore, an estimation of the density of states at the Fermi energy $N(E_F)$ using the $Bc2(0)$ and normal state resistivity data suggests that $N(E_F)$ is decreased too, being consistent with theoretical calculations.

## IV. ACKNOWLEDGEMENTS

The work is supported by the National Center for Research and Development on Superconductivity and the Ministry of Science and Technology (NKBRSF-G19990646). One of us (JQL) acknowledges the support of 'Hundred of Talents' program organized by the Chinese Academy of Sciences, P.R. China.

## References:

[*] Corresponding author, E-mail: Dzheng@ssc.iphy.ac.cn
[1] J. Nagamatsu, N. Nakagawa, T. Muranaka, Y. Zenitani, and J. Akimitsu, *Nature* **410**, 63 (2001).
[2] M. E. Jones and R. E. Marsh, *J. Amer. Chem. Soc.* **76**, 1434 (1954).
[3] J. Kortus, I. I. Mazin, K. D. Belashchenko, V. P. Antropovz, and L. L. Boyery, *cond-mat/0101446* (2001).
[4] S. L. Bud'ko, G. Lapertot, C. Petrovic, C. E. Cunningham, N. Anderson, and P. C. Canfield, *Phys. Rev. Lett.* **86**, 1877(2001).
[5] H. Schmdit, J. F. Zasadzinski, K. E. Gray, and D. G. Hinks, *cond-mat/0102389* (2001).
[6] A. Sharoni, Israel Felner, and Oded Millo, *cond-mat/0102325* (2001).
[7] R. K. Kremer, B. J. Gibson, and K. Ahn, *cond-mat/0102432* (2001).
[8] T. Takahashi, T. Sato, S. Souma, T. Muranaka, and J. Akimitsu, *cond-mat/0103079* (2001).
[9] T. Yidirim, O. Gulseren, J. W. Lynn, C. M. Brown, T. J. Udovic, H. Z. Qing, N. Rogado, K. A. Regan, M. A. Hayward, J. S. Slusky, T. He, M. K. Haas, P. Khalifah, K. Inumaru, and R. J. Cava, *cond-mat/0103469* (2001).
[10] J. E. Hirsch, *cond-mat/0102115* (2001).
[11] J. E. Hirsch and F. Marsiglio, *cond-mat/0102479* (2001).
[12] W. N. Kang, C. U. Jung, H. P. Kim, Min-Seok Park, S. Y. Lee, Hyeong-Jin Kim, Eun-Mi Choi, Kyung Hee Kim, Mun-Seog Kim, and Sung-Ik Lee, *ond-mat/0102313* (2001).
[13] J. S. Slusky, N. Rogada, K. A. Regan, M. A. Hayward, P. Khalifah, T. He, K. Lnumaru, S. M. Loureiro, M. K. Hass, H. W. Zandbergen, and R. J. Cava, *Nature* **410**, 343 (2001).
[14] B. Lorenz, R. L. Meng, Y. Y. Xue, and C. W. Chu. *cond-mat/0104041* (2001).

[15] D. K. Finnemore, J. E. Ostenson, S. L. Bud'ko, G. Lapertot, and P. C. Canfield, *Phys. Rev. Lett*. **86**, 2420 (2001).
[16] A. L. Fetter and P. C. Hohenberg, in *Superconducitivity* edited by R. D. Parks,(Marcel, Dekker, New York, 1969) P.817.
[17] B. Miihlschlegel, *Z.Phys*. **155**, 313 (1959).
[18] N. R. Werthamer, E. Helfand, and P. C. Hohenberg, *Phys. Rev*. **147**, 288 (1966).
[19] K. Maki, *Phys. Rev*. **148**, 362 (1966).
[20] G. Eilenberger, *Phys. Rev*. **153**, 584 (1967).
[21] T. E. Faber and A. B. Pippard, *Proc. Roy. Soc*. **A231**, 336 (1955).
[22] P. C. Canfield, D. K. Finnemore, S. L. Bud'ko, J. E. Ostenson, G. Lapertot, C. E. Cunningham, and C. Petrovic, *Phys. Rev. Lett,* **86**, 2423 (2001).
[23] W. N. Kang, Hyeong-Jin Kim, Eun-Mi Choi, C. U. Jung, and Sung-Ik Lee, *Science*, **292**, 1521 (2001).
[24] W. N. Kang, Hyeong-Jin Kim, Eun-Mi Choi, Kijoon H. P. Kim, and Sung-Ik Lee, *cond-mat/0105024 (*2001).
[25] C. B. Eom, M. K. Lee, J. H. Choi, L. J. Belenky, X. Song, L. D. Cooley, M. T. Naus, S. Patnaik, J. Jiang, M. Rikel, A. Polyanskii, A. Gurevich, X. Y. Cai, S. D. Bu, S. E. Babcock, E. E. Hellstrom, D. C. Larbalestier, N. Rogado, K. A. Regan,M. A. Hayward, T. He, J. S. Slusky, K. Inumaru, M. K. Hass, and R. J. Cava, *Nature* **441**, 558 (2001).
[26] S. Patnaik, L. D. Cooley, A. Gurevich, A. A. Polyanskii, J. Jiang, X. Y. Cai, A. A. Squitieri, M. T. Naus, M. K. Lee, J. H. Choi, L. Belenky, S. D. Bu, J. Letteri, X. Song, D. G. Schlom, S. E. Babcock, C. B. Eom, E. E. Hellstrom, and D. C. Larbalestier, *cond-mat/0104562* (2001).
[27] D. R. Armstrong and P. G. Perkins, *J. Chem. Soc. Faraday. Trans*. **275**, 13 (1979)
[28] A. L. Ivanovskii and I. Medvedeva, *Russ. J. Inorg. Chem.* **45,** 1234 (2000).
[29] T. P. Orlando, E. J. McNiff, Jr, S. Foner, and M. R. Beasley, *Phys. Rev. B* **19**, 4545 (1979).
[30] J. Q. Li, L. Li, F. M. Liu, C. Dong, J. Y. Xiang, and Z. X. Zhao, *cond-mat/0104320* (2001).

Figure caption:

FIG. 1: partial representation of X-ray diffraction patterns for a series of $Mg_{1-x}Al_xB_2$ samples.

FIG. 2: The lattice parameter *a* and *c* as a function of Al concentration. In the shadowed region, the *(002)* reflection peak is broad, indicating the partial collapse of the separation between boron layers.

FIG. 3: Zero-field-cooled (ZFC) magnetization as a function of temperature for different Al doping level. The inset is a magnified part of one of the transition curve to show the definition of *Tc*(*bulk*) and *Tc*(*onset*).

FIG. 4: Variation of the onset superconducting transition temperature *Tc*(*onset*) and bulk superconducting transition temperature *Tc*(*bulk*) with Al doping level. Due to the rounded curvature between *Tc*(*bulk*) and *Tc*(*onset*) there is a large error bar associated with *Tc*(*onset*). The curve of *Tc*(*onset*) (open circles) defined in Fig.3 might be regarded as the lower bound while the dashed line is the upper bound of *Tc*(*onset*).

FIG. 5: Temperature dependence of the upper critical field $Bc2(0)$ for the *x*=0 and 0.07 samples. The inset shows the FC and ZFC magnetization data as a function of temperature measured at 0.5 T for the *x*=0.07 sample. The superconducting transition temperature *Tc* and the irreversibility temperature $T_{irr}$ are indicated in the graph.

FIG. 6: TEM diffraction patterns for the two $Mg_{1-x}Al_xB_2$ samples. a) *x*=0, b) *x*=0.25.

FIG. 7: A dark-field image of a grain for the *x*=0.25 sample.

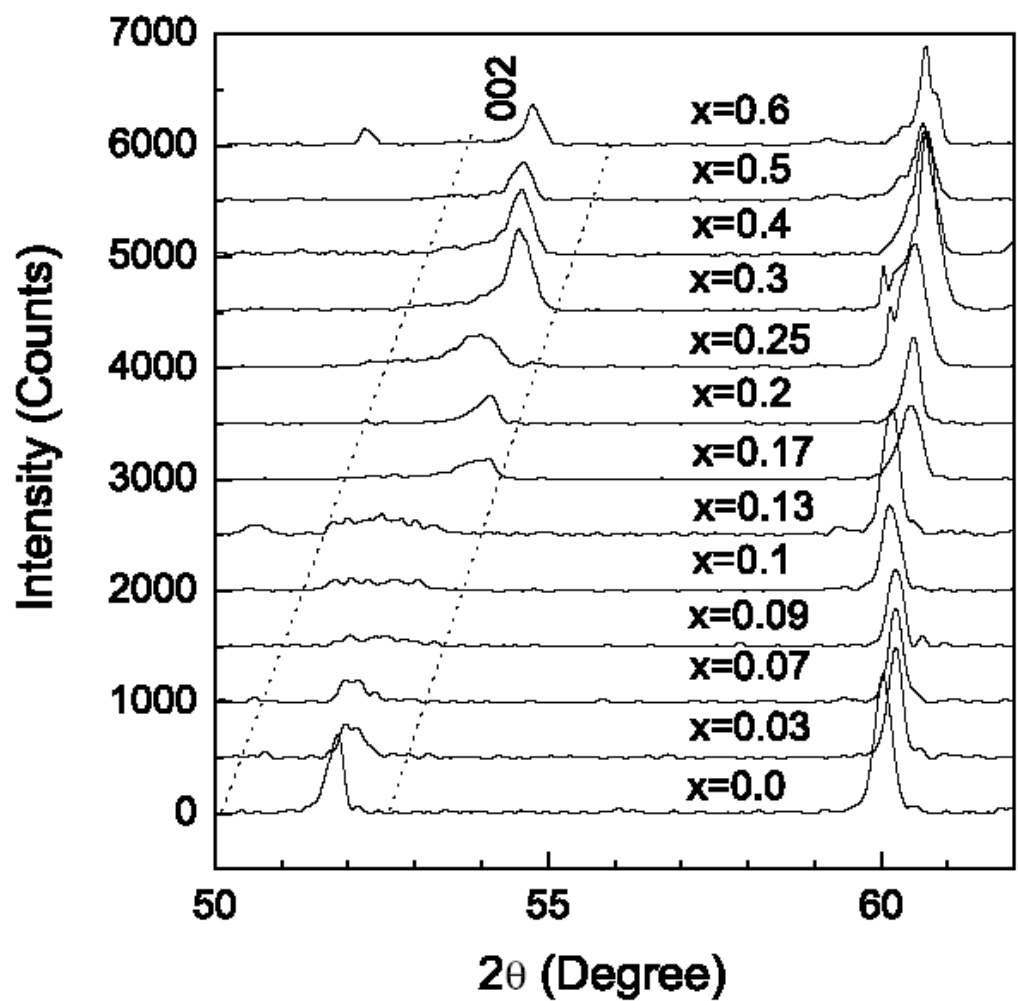

Xiang et al     FIG.1

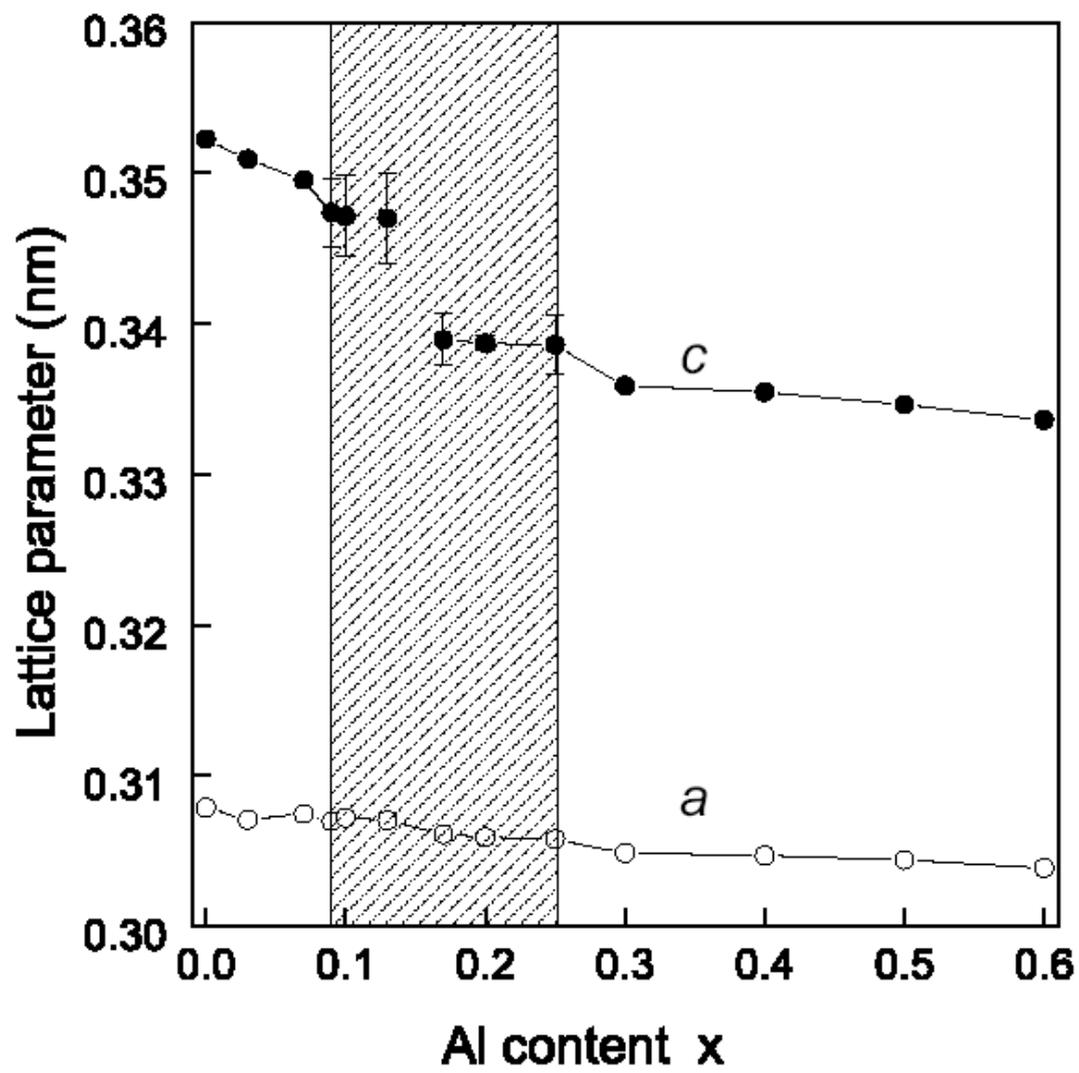

Xiang et al                FIG.2

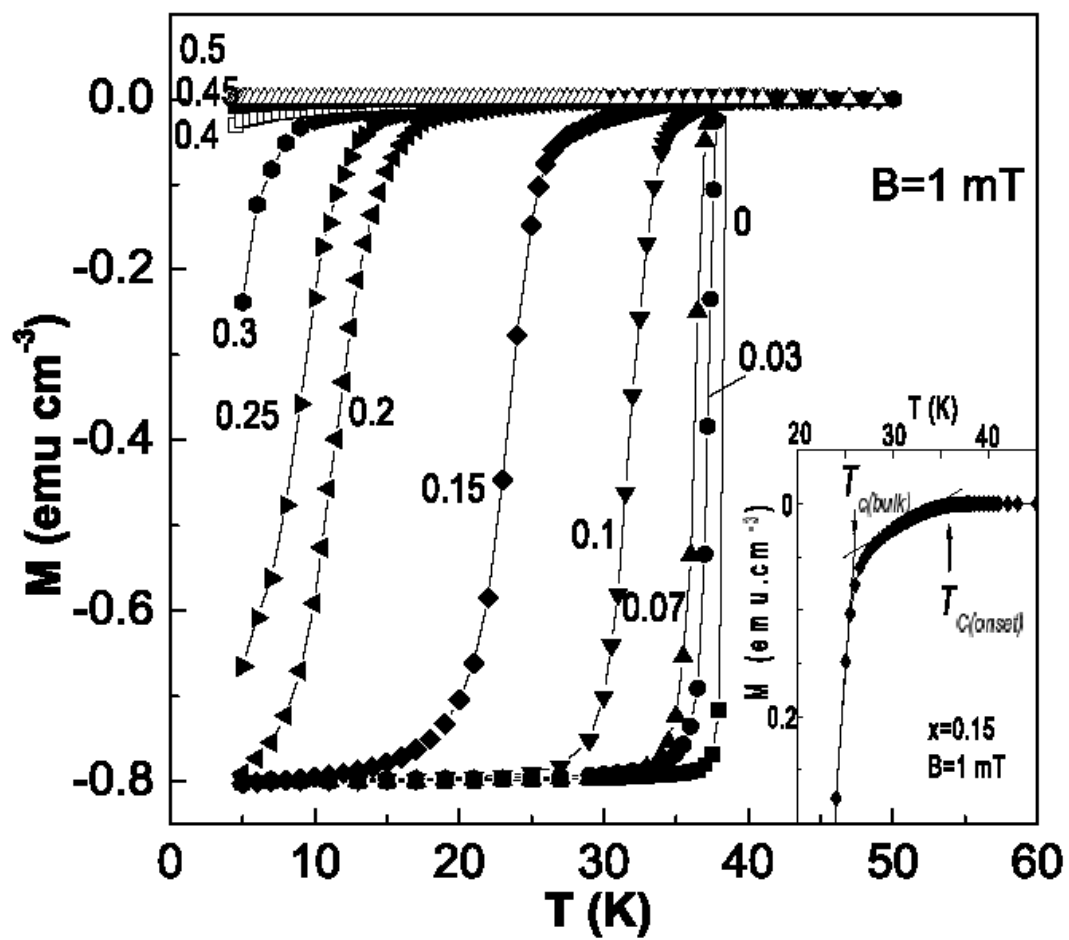

Xiang at al      FIG.3

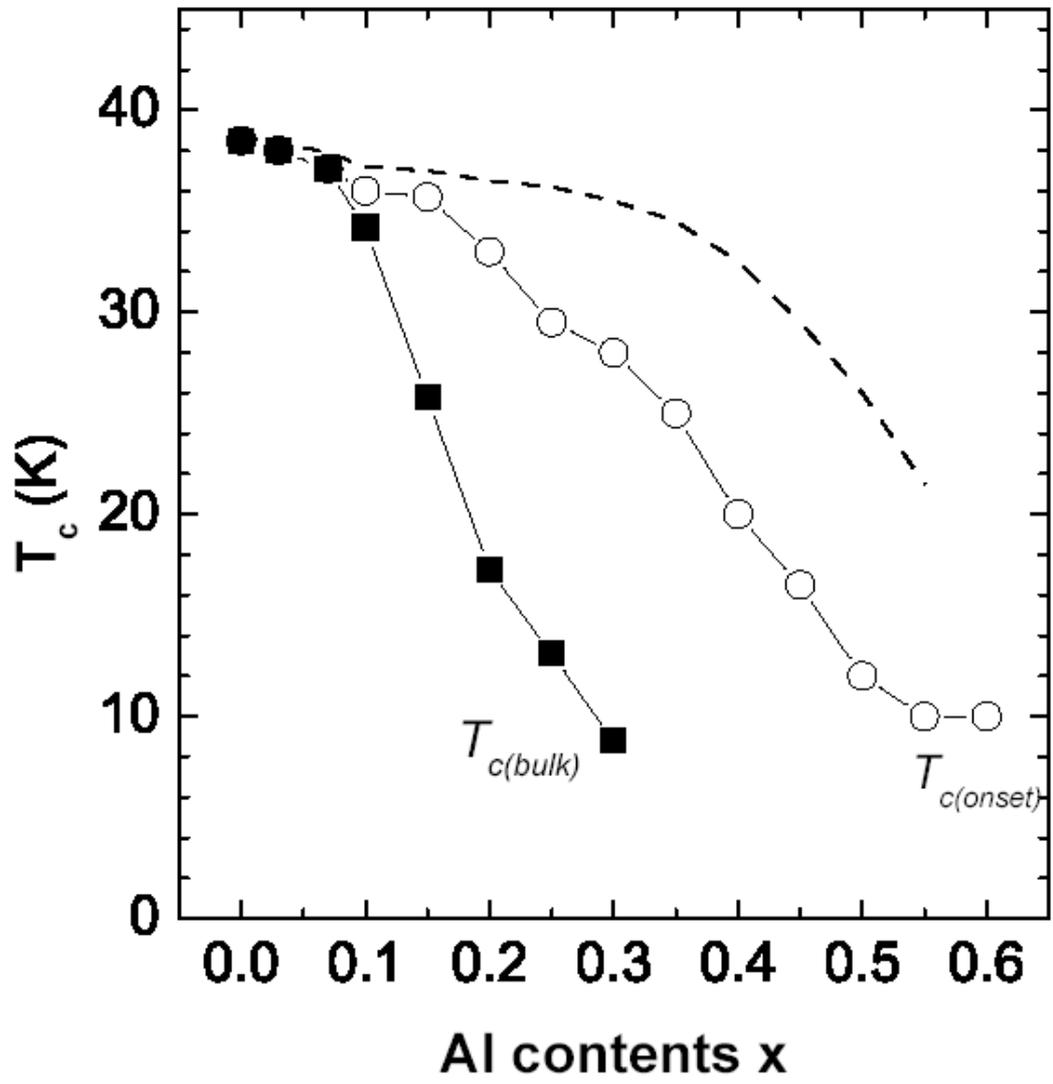

Xiang et al         FIG.4

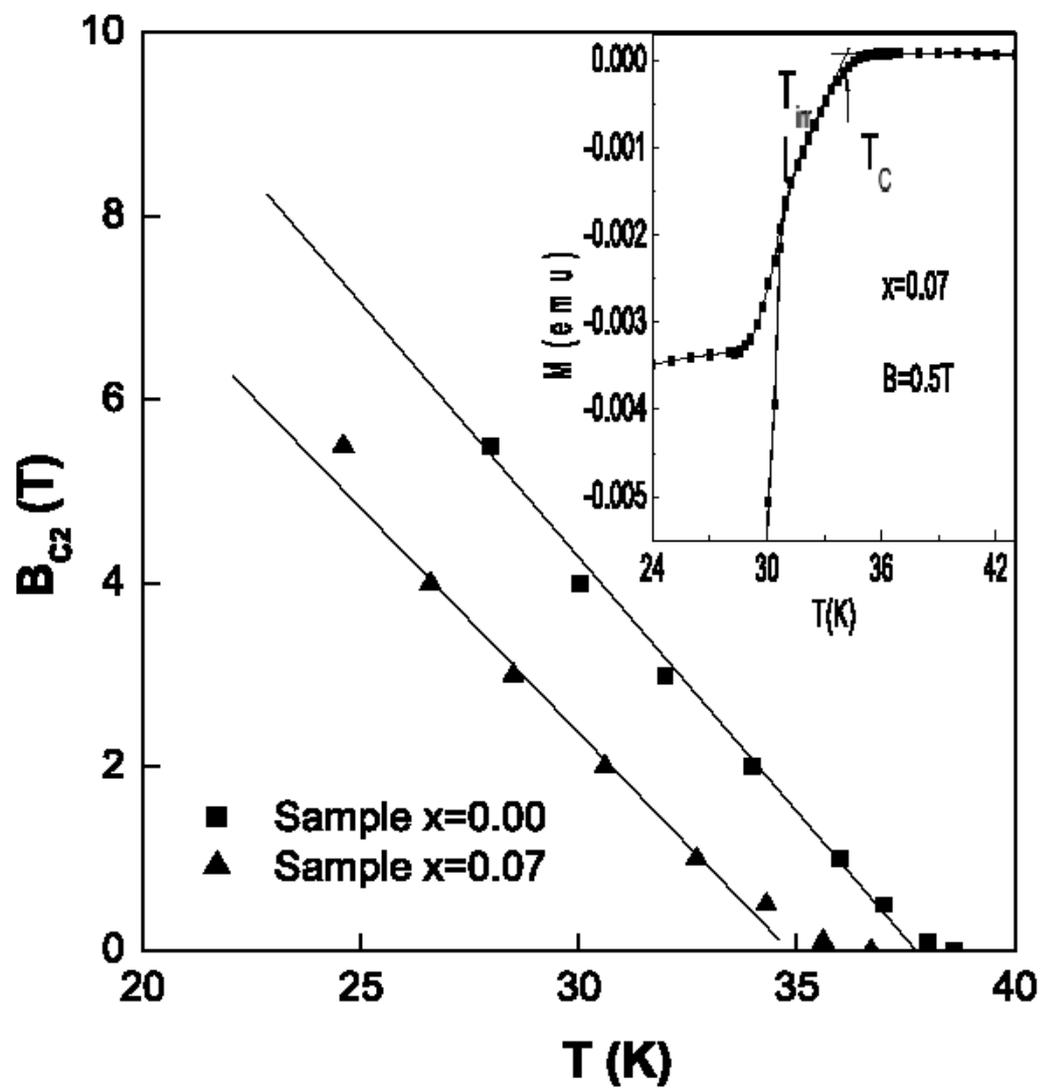

Xiang et al     FIG.5

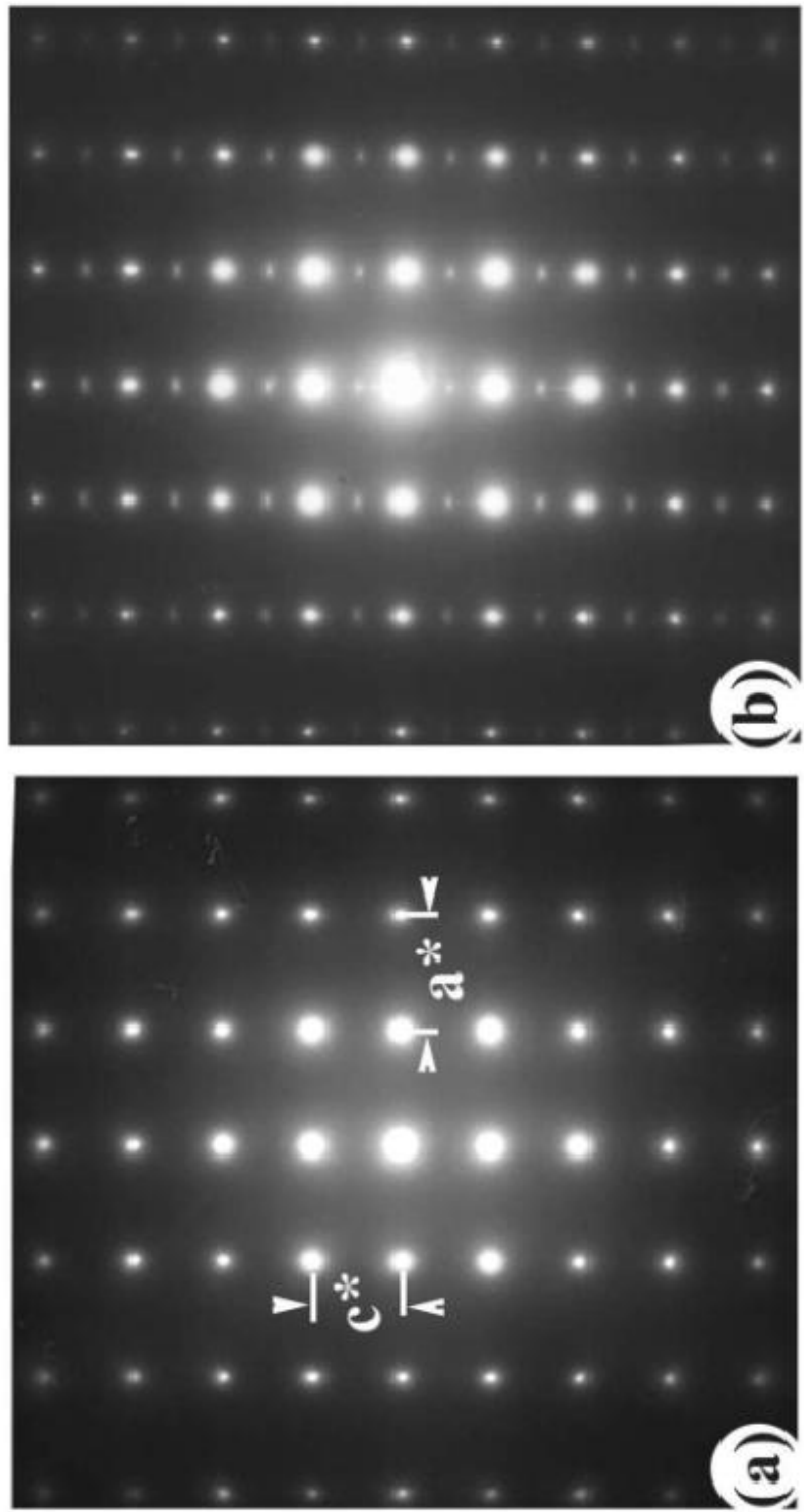

Xiang et al     FIG.6

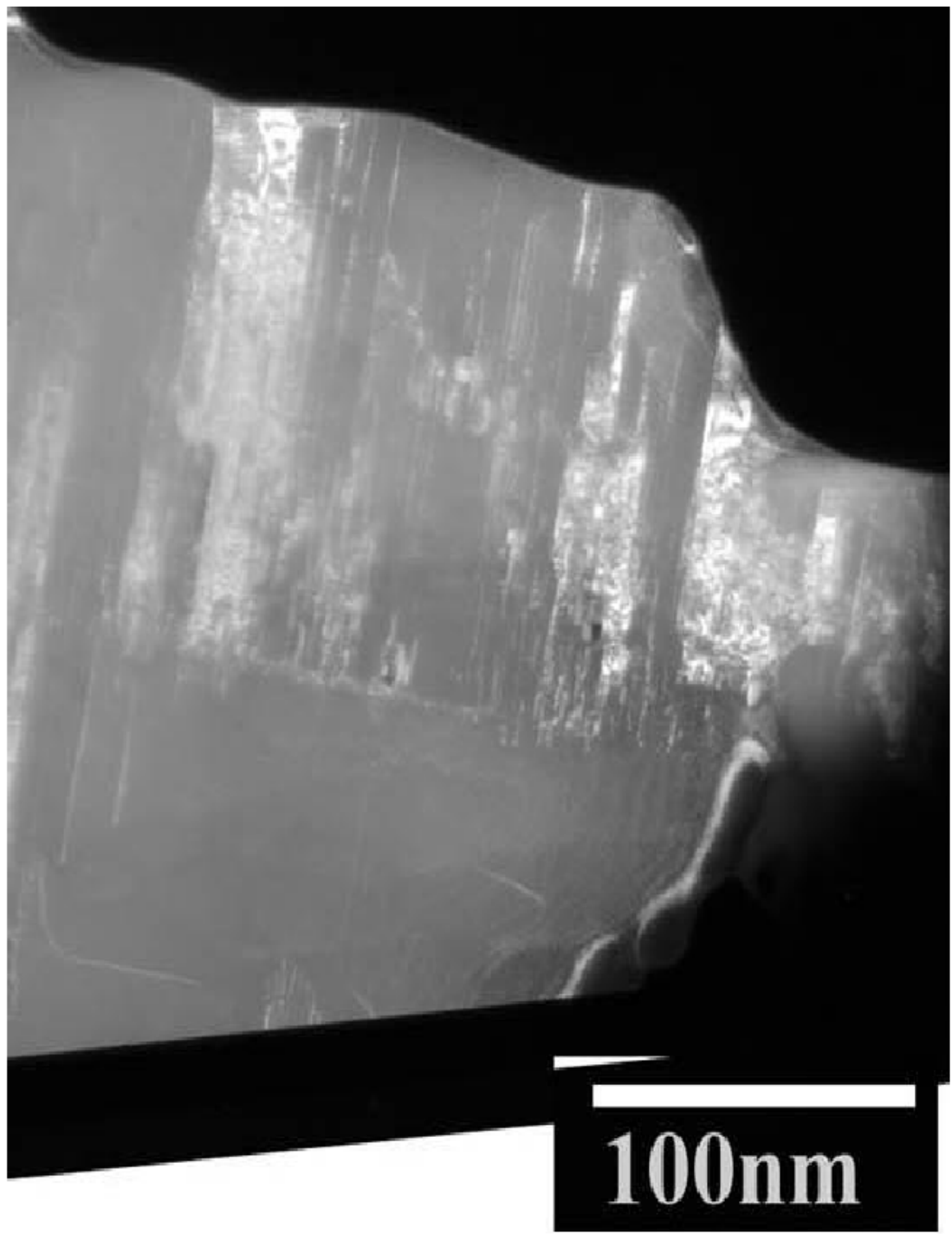

**Xiang et al**     **FIG.7**